\documentclass{IEEEcsmag}

\usepackage[colorlinks,urlcolor=blue,linkcolor=blue,citecolor=blue]{hyperref}
\expandafter\def\expandafter\UrlBreaks\expandafter{\UrlBreaks\do\/\do\*\do\-\do\~\do\'\do\"\do\-}
\usepackage{upmath}

\usepackage{amsmath,amssymb,amsfonts}
\usepackage{algorithmic}
\usepackage{graphicx}
\usepackage{textcomp}
\usepackage{xcolor}
\usepackage{subcaption}
\usepackage{todonotes}

\usepackage{booktabs}
\usepackage[braket,qm]{qcircuit}

\usepackage{eso-pic}

\jvol{XX}
\jnum{XX}
\paper{8}
\jmonth{October}
\jname{IEEE CG\&A}
\jtitle{Quantum Machine Learning Playground}
\pubyear{2024}

\begin{document}
\AddToShipoutPictureFG{
  \put(100,-65){%
    \parbox[b]{\paperwidth}{%
      \fontsize{5pt}{10pt}\selectfont
      \centering
      \textcolor{black}{© 2024 IEEE. Personal use is permitted, but republication/redistribution requires IEEE permission. See \texttt{https://www.ieee.org/publications/rights/index.html} for more information.}%
    }
  }
  \put(170,690){%
    \parbox[b]{0.575\paperwidth}{%
      \fontsize{5pt}{10pt}\selectfont
      \centering
      \textcolor{black}{This article has been accepted for publication in IEEE Computer Graphics and Applications. This is the author's version which has not been fully edited and content may change prior to final publication. Citation information: DOI 10.1109/MCG.2024.3456288}%
    }
  }
}
\sptitle{Theme Article: Special Issue on Quantum Visual Computing}

\title{Quantum Machine Learning Playground}

\author{Pascal Debus\IEEEauthorrefmark{1}, Sebastian Issel\IEEEauthorrefmark{1}, Kilian Tscharke\IEEEauthorrefmark{1}\\
\IEEEauthorrefmark{1}Fraunhofer Institute for Applied and Integrated Security, Garching near Munich, 85748, Germany}
%\affil{Fraunhofer Institute for Applied and Integrated Security, Garching near Munich, 85748, Germany}

% \author{Sebastian Issel}
% \affil{Fraunhofer Institute for Applied and Integrated Security, Garching near Munich, 85748, Germany}

% \author{Kilian Tscharke}
% \affil{Fraunhofer Institute for Applied and Integrated Security, Garching near Munich, 85748, Germany}

\markboth{Quantum Visual Computing}{Quantum Visual Computing}

\begin{abstract}\looseness-1 This article introduces an innovative interactive visualization tool designed to demystify quantum machine learning (QML) algorithms. Our work is inspired by the success of classical machine learning visualization tools, such as TensorFlow Playground, and aims to bridge the gap in visualization resources specifically for the field of QML. The article includes a comprehensive overview of relevant visualization metaphors from both quantum computing and classical machine learning, the development of an algorithm visualization concept, and the design of a concrete implementation as an interactive web application. By combining common visualization metaphors for the so-called data re-uploading universal quantum classifier as a representative QML model, this article aims to lower the entry barrier to quantum computing and encourage further innovation in the field. The accompanying interactive application is a proposal for the first version of a quantum machine learning playground for learning and exploring QML models.
\end{abstract}

\maketitle

\chapteri{Q}uantum computing has the potential to bring disruptive changes in a variety of fields such as cryptography, operations research, chemistry, and artificial intelligence. As a result, finding out which quantum algorithm might result in a so-called \emph{quantum advantage} for a particular application is a very active field of research that requires knowledge in both quantum computing and the application area.

However, the abstract and intangible nature of quantum states and operations poses a significant barrier to the understanding of quantum algorithms, especially for newcomers to the field. Visualization plays a critical role in understanding and teaching quantum computing concepts. While there are resources available for learning quantum computing, a gap exists in tools specifically designed for interactive learning and exploration of quantum machine learning algorithms. Tools like IBM's Quantum Experience and Quirk provide platforms for visualizing quantum circuits and their execution. However, these tools often focus on circuit representation and lack interactive features for exploring the dynamics of quantum algorithms.

Our work focuses on the development of an interactive visualization application for a representative QML model, the data re-uploading universal quantum classifier \cite{perez-salinas_data_2020}. Our approach is inspired by the success and pedagogical approach of Google's TensorFlow Playground \cite{smilkov_direct-manipulation_nodate} in the domain of classical machine learning. To address this challenge, our proposed interactive visualization tool aims to demystify the inner workings of a quantum classifier by providing a hands-on, visual exploration environment. 
To that end, we also introduce \emph{Q-simplex}, a novel visualization method for two-qubit states.
Our application enables users to intuitively understand how the data is processed, how the quantum gates affect the state, and how the algorithm iteratively adjusts its parameters to improve classification accuracy. This visual and interactive approach not only aids in educational efforts by making quantum computing concepts more tangible but also serves as a powerful research tool for exploring QML models.

\subsection{Related Work}
To the best of our knowledge, there is no related work specifically focusing on the visualization of quantum machine learning yet. Most publications consider either general quantum state space visualization \cite{lamy_dynamic_2019, ruan_venus_2023}, visualization of quantum circuits \cite{wen_quantivine_2023, williams_qcvis_2021}, or the visualization of a few very well-known quantum algorithms \cite{karafyllidis_visualization_2003}, such as Shor's or Grover's. Furthermore, \cite{bethel_quantum_2023,tao_shorvis_2017} provide an overview of the various visualization techniques.

\subsection{Structure}
The article is structured as follows: In the next section, we provide the necessary theoretical background of quantum computing, neural networks, and quantum machine learning. Next, we give an overview of the most relevant visualization metaphors in quantum computing and classical machine learning. These are the basics for the following section on our algorithm visualization concept whose concrete implementation as an interactive web application is subsequently discussed in the section on implementation details. Finally, we provide some directions for future research and a conclusion.

%\vadjust{\pagebreak} 

\section{BACKGROUND}
%\section{Background}
The field of quantum computing represents a fundamental shift from classical computing paradigms, leveraging the principles of quantum mechanics to perform computations. This section provides an overview of the essential concepts in quantum computing and introduces the basic principles of quantum machine learning (QML) necessary to understand the data re-uploading universal quantum classifier model.

\subsection{Quantum Computing}
At the heart of quantum computing is the concept of the qubit, short for "quantum bit".
Unlike classical bits, which can be either 0 or 1, qubits can exist in a state of superposition.
This property allows quantum computers to process both possibilities simultaneously, offering potential speedups for certain computational problems.

\subsubsection{Single Qubit States}
A qubit's state is represented as a linear combination of the basis states \(\ket{0}\) and \(\ket{1}\):
\begin{equation}
\ket{\psi} = \alpha\ket{0}+\beta\ket{1} = \alpha \begin{pmatrix} 1 \\ 0 \end{pmatrix} + \beta \begin{pmatrix} 0 \\ 1 \end{pmatrix}
\label{eq:qubit_state}
\end{equation}
where $\alpha$ and $\beta$ are \emph{complex} amplitudes $\alpha,\beta\in\mathbb{C}$. The probabilities of measuring the qubit in either state are given by \(|\alpha|^2\) and \(|\beta|^2\) respectively, resulting in the normalization constraint \(|\alpha|^2 + |\beta|^2 = 1\).
Consequently, a state can also be represented as a two-dimensional, complex vector $(\alpha, \beta)\in\mathbb{C}^2$ with unit norm.

\subsubsection{Quantum Gates}
Quantum gates manipulate the state of qubits and are the building blocks of quantum algorithms.
These gates are represented by unitary matrices, ensuring that the quantum state remains normalized (the total probability remains 1). As such, they can be generally interpreted as rotations in a complex vector space.
This becomes clear when looking at the Bloch sphere visualization in the next section.

Common elementary gates include the rotation gates $R_X$, $R_Y$, and $R_Z$, which rotate the qubit around the x-, y-, and z-axes of the Bloch sphere.
Then there are the Pauli Matrices $X$, $Y$, and $Z$, as special cases (or generators) of these rotations.
Since unitary matrices are generalized rotations, every unitary matrix can be expressed as a series of rotations around two axes.
The action of a gate on a qubit can be computed using matrix-vector multiplication, e.g., the Pauli $X$ gate acts on a basis state of a qubit like a classical NOT gate, by inverting the bits.
\begin{align*}
    X\; &\widehat{=} \begin{pmatrix}
    0 & 1\\ 1 & 0    
    \end{pmatrix} \\ 
    X\ket{\psi} &= X(\alpha\ket{0}+\beta\ket{1}) \\
    &\widehat{=} \begin{pmatrix}
    0 & 1\\ 1 & 0    
    \end{pmatrix}
    \begin{pmatrix}
    \alpha \\ \beta    
    \end{pmatrix}=
    \begin{pmatrix}
    \beta \\ \alpha    
    \end{pmatrix} \\
    &\widehat{=}\; \beta\ket{0}+\alpha\ket{1}
\end{align*}
Consequently, when applied to the general superposition state described in Equation~\ref{eq:qubit_state} both basis states will be flipped which results in swapping the coefficients $\alpha$ and $\beta$.

\subsubsection{Multi-Qubit States}
Two qubits can exist in a superposition of four states:
$\ket{00}$, $\ket{01}$, $\ket{10}$, and $\ket{11}$.
The state $\ket{\psi^{(2)}}$ of a two-qubit system can be described as:

\begin{equation}
\ket{\psi^{(2)}} = a\ket{00} + b\ket{01} + c\ket{10} + d\ket{11}    
\label{eq:twoqubit_state}
\end{equation}

where \(a\), \(b\), \(c\), and \(d\) are complex numbers encoding the probability amplitudes of each state, with the constraint that \(|a|^2 + |b|^2 + |c|^2 + |d|^2 = 1\).
When dealing with multiple qubits, the system's state space grows exponentially.
For an \(n\)-qubit system, the state vector $\ket{\psi^{(n)}}$ lives in a $2^n$-dimensional Hilbert space described by $2^n$ complex amplitudes $\alpha_k$:
\begin{equation}
    \ket{\psi^{(n)}} = \sum_{k=0}^{2^n-1}\alpha_k\ket{k},
    \label{eq:nqubit_state}
\end{equation}
where $\ket{k}$ corresponds to the binary expansion $\ket{b_{n-1}^{(k)}\cdots b_{1}^{(k)} b_{0}^{(k)}}$ of  $k$. 

\subsubsection{Multi-qubit gates}
Multi-qubit gates are necessary to generate entangled states, a resource for quantum communication and computation. The significance of entangled states is that the state of one of its constituent qubits cannot be described independently of another of its qubits (nonseparability), leading to correlations that are crucial for quantum computing's power. The Bell states, such as $\ket{\Phi_+} := \frac{1}{\sqrt{2}}(\ket{00} + \ket{11})$, are famous examples of (maximally) entangled states.The notion of \textit{maximally} entangled states implies that entanglement can be quantified beyond being just separable or not separable. There are multiple measures that quantify entanglement, however, in this short introduction, we will only mention the \emph{concurrence} $C$. For the case of the two-qubit state defined in Equation \eqref{eq:twoqubit_state} it is given by
\begin{equation*}
    C:=2|ad-bc|\leq 1.
\end{equation*}
With this metric, a maximally entangled state has $C=1$ while a separable state has $C=0$.

As in the case of single-qubit gates, multi-qubit gates can be represented by unitary matrices acting on vectors in a complex vector space. 
Furthermore, multi-qubit gates acting on $n$ qubits can always be decomposed into a series of gates acting on only two or one qubits. Hence, as in classical computing, there exist sets of gates that are universal in the sense that any unitary gate can be written as a composition of gates from such a set. 

A typical member of such a universal gate set is the CNOT Gate (Controlled-NOT): A two-qubit gate where the state of the second qubit, the target, is flipped if (and only if) the first qubit, the control, is in state $\ket{1}$.
Mathematically, it can be represented as:
\begin{align*}
\operatorname{CNOT} &= \begin{pmatrix}
1 & 0 & 0 & 0 \\
0 & 1 & 0 & 0 \\
0 & 0 & 0 & 1 \\
0 & 0 & 1 & 0 \\
\end{pmatrix}\\
\operatorname{CNOT}\ket{\psi^{(2)}}&=a\ket{00} + b\ket{01} + d\ket{10} + c\ket{11}
\end{align*}

\subsubsection{Measurements}
Finally, we tend to the measurement process which is the last missing piece in this minimal introduction to quantum computing. When a measurement is made, the superposition collapses to a single state, with the outcome determined by the probabilities associated with each state. Here, we will only consider simple measurements with respect to the so-called computational basis $\lbrace\ket{0}=(1\;0)^T,\, \ket{1}=(0\;1)^T\rbrace$. With the measurement of the state defined in Equation~\eqref{eq:nqubit_state} it will collapse into one of the states $\ket{k}$ with probability 
$P_k = \lvert \alpha_k \rvert^2$.

The general implications for the design of any quantum algorithm are that the solution of a problem must be encoded in a basis state $\ket{k}$ that has a high probability of being measured. For classification tasks in machine learning, for example, one can encode the different classes using the basis $C_0\mapsto\ket{00}$, $C_1\mapsto\ket{01}$, $C_2\mapsto\ket{10}$, $C_3\mapsto\ket{11}$ and for a data sample belonging to class $C_0$, there must be a high probability of measuring $\ket{00}$.

\subsection{Machine Learning}
In the following subsection, we provide a brief mathematical introduction to feed-forward neural networks in classical machine learning. This will provide a foundation for understanding the parallels and contrasts with quantum machine learning models, particularly the data re-uploading quantum classifier.

Feed-forward neural networks (FNNs) are the simplest type of artificial neural network architecture, wherein connections between the nodes do not form cycles. A FNN consists of multiple layers, including an input layer, one or more hidden layers, and an output layer. Each layer is made up of nodes, or neurons, which apply an activation function to the weighted sum of inputs from the previous layer plus a bias.

\begin{figure*}[ht]
\centerline{\includegraphics[width=0.85\linewidth]{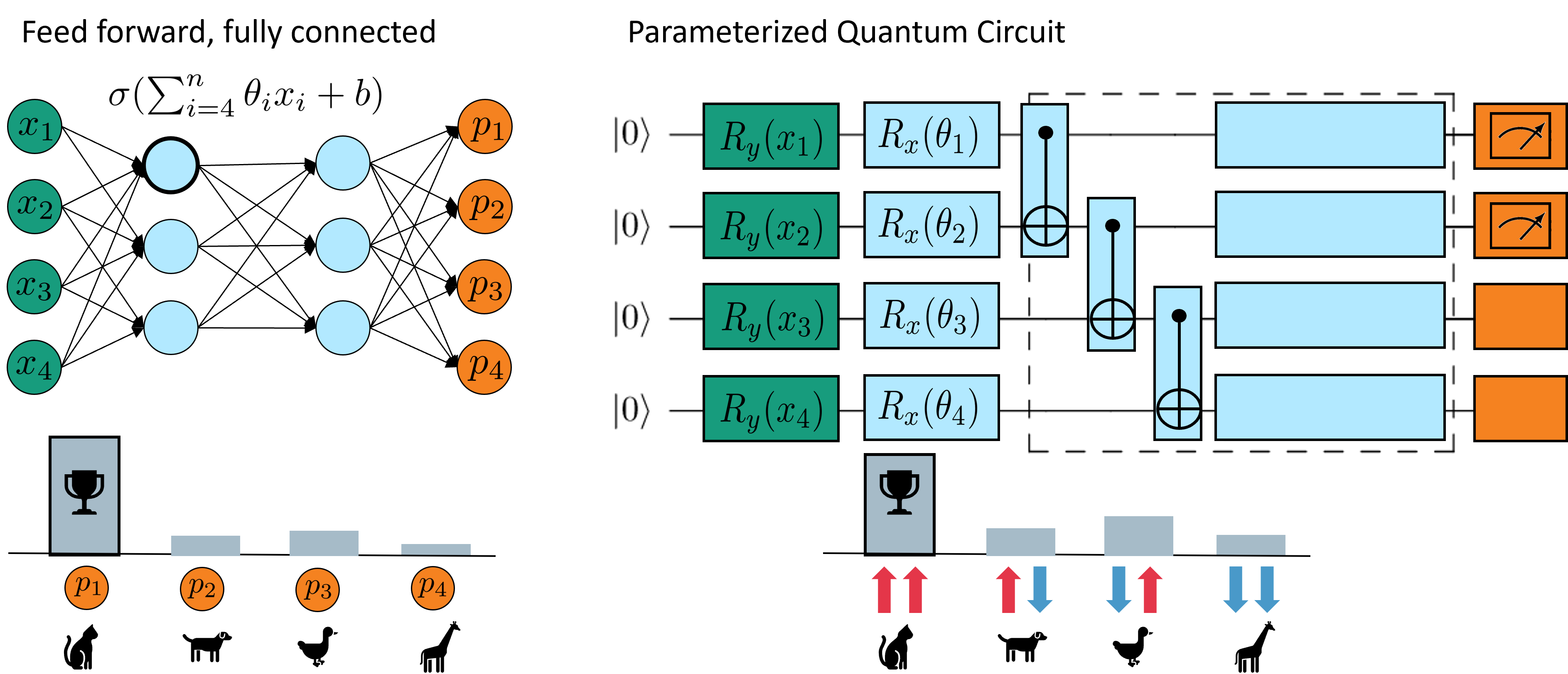}}
\caption{Comparison of a feed-forward neural network with a QML model based on a parameterized quantum circuit. The histograms at the bottom show the probabilities of the four different classes (e.g., cat, dog, duck, giraffe). For the classical model, those are obtained by normalizing the outputs of the last layer while for the quantum model, they are obtained from multiple measurements of the circuit and assigning classes to basis states.}\vspace*{-5pt}
\label{fig:ffnn_qnn}
\end{figure*}

The output \(y\) of a neuron can be mathematically represented as:
\begin{equation*}
   y = \sigma\left(\sum_{i=1}^{n} \theta_i x_i + b\right) 
\end{equation*}
where \(x_i\) are the inputs, \(\theta_i\) are the weights, \(b\) is the bias, and \(\sigma\) is an activation function. This is also illustrated in Figure~\ref{fig:ffnn_qnn} for $n=4$ inputs to the highlighted neuron.

The network learns by adjusting the weights and biases to minimize the difference between the actual output and the target output. This process is typically achieved through backpropagation and an optimization algorithm like gradient descent.

\subsection{Quantum Machine Learning}
Quantum Machine Learning (QML) is an emerging field that combines quantum computing with machine learning algorithms. For an extensive overview consider, for example, \cite{schuld_machine_2021}.
QML  leverages quantum algorithms to perform machine learning tasks, potentially offering significant speedups and efficiency improvements over classical algorithms for certain problems. 
One approach in QML is the use of parameterized quantum circuits (PQCs), which are structurally similar to neural networks in classical machine learning but operate under quantum mechanical principles.  Here, a quantum circuit is designed to classify data by iteratively adjusting its parameters based on the input data.
This process can be seen as an analog to the way FNNs adjust their weights and biases during training.

In a simplistic model, a quantum classifier can be represented by a quantum circuit that applies a series of parameterized gates \(U(\theta)\) on an initial state $\ket{0}^n=\ket{00\ldots0}$, where \(n\) is the number of qubits, and \(\theta\) represents the parameters of the circuit. The output of the circuit is then measured to classify the input data.
A loss function is then used to compute gradients and adjust \(\theta\) such that the measurement outcomes correspond to the correct classification of the input data.

Mathematically, the action of a parameterized quantum gate on a qubit can be represented as:
\[ |\psi(\mathbf{\theta})\rangle = U(\mathbf{\theta})|0\rangle \]
where \(|\psi(\mathbf{\theta})\rangle\) is the state of the qubit after the application of \(U(\mathbf{\theta})\), and \(|0\rangle\) is the initial state of the qubit. The unitary  \(U(\mathbf{\theta})\) is typically a short-hand notation for a series of 1- and 2-qubit gates that are grouped into layers and make up the architecture of the quantum model.

\begin{figure}[ht]
\[
\Qcircuit @C=0.85em @R=.9em  @!R{
 & & \mbox{$L(1)$} & & & & & & \mbox{$L(N)$} & &\\
\lstick{|0\rangle} & \gate{U\left(\mathbf{x}\right)} & \qw & \gate{U(\mathbf{\theta}_{1})} & \qw & \cdots & &  \gate{U\left(\mathbf{x}\right)} & \qw & \gate{U(\mathbf{\theta}_{N})} & \qw & \meter
\gategroup{2}{2}{2}{4}{.7em}{--}
\gategroup{2}{8}{2}{10}{.7em}{--}
}
\]
\\
\[
\Qcircuit @C=0.9em @R=.9em  @!R{
% & \mbox{$L(1)$} & & & & \mbox{$L(N)$} & \\
\lstick{|0\rangle} & \qw & \qw & \gate{U\left(\mathbf{\theta}_{1},\mathbf{x}\right)} & \qw & \qw & \cdots & & \qw & \qw &  \gate{U\left(\mathbf{\theta}_{N},\mathbf{x}\right)} & \qw & \qw & \meter
\gategroup{1}{4}{1}{4}{.7em}{--}
\gategroup{1}{11}{1}{11}{.7em}{--}
}
\]
\caption{Quantum Circuit for the data re-uploading classifier. The upper circuit shows the variant with separate gates for encoding and trainable weights, and the lower circuit the compact variant with just one gate per layer. Adapted from \cite{perez-salinas_data_2020}.}
\label{fig:data_reuploading}
\vspace*{-5pt}
\end{figure}
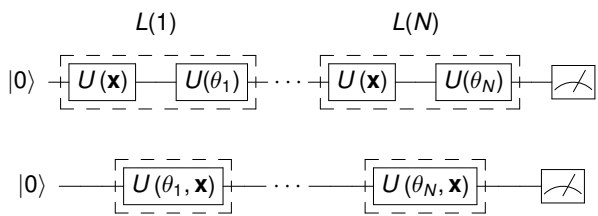

The classifier's performance and its ability to generalize well to unseen data depend on the design of the quantum circuit (the arrangement and types of gates used) and the optimization of the parameters $\mathbf{\theta}$.

\begin{figure}[ht]
\centerline{\includegraphics[width=\linewidth]{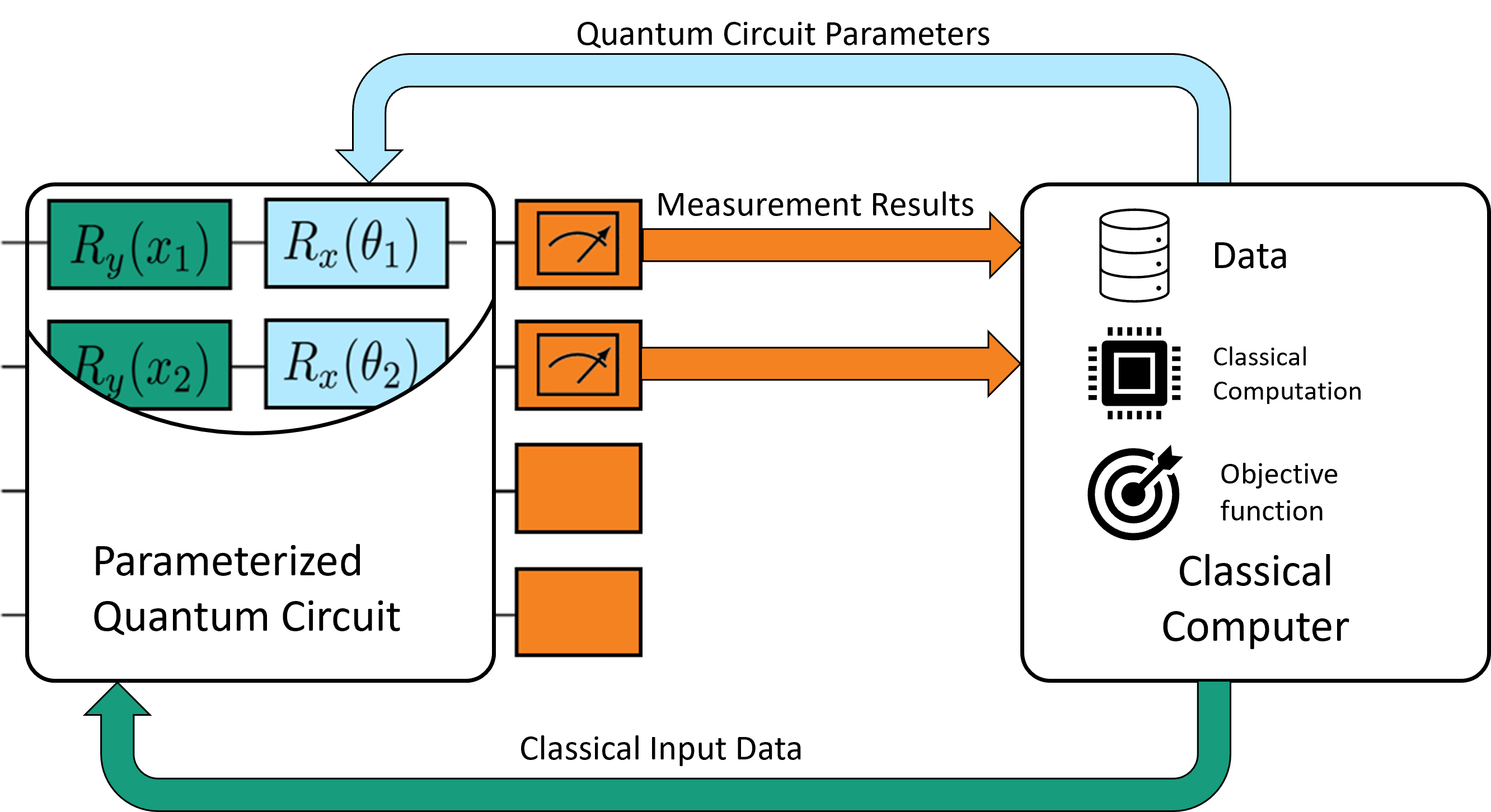}}
\caption{Illustration of a training cycle for parameterized quantum circuits.}
\vspace*{-5pt}
\end{figure}

\subsubsection{Data Re-uploading Quantum Classifier}
The data re-uploading quantum classifier \cite{perez-salinas_data_2020} is a prominent example of a QML model that utilizes a PQC where data is repeatedly encoded into the quantum state, interspersed with parameterized gates for processing.

The model can be represented as follows:
\begin{equation*}
U(\mathbf{\theta}, \mathbf{x}) = L(N)\cdots L(2)L(1)   
\end{equation*}
where $L(k)$ is either a product of  a  data-encoding unitary \(U(\mathbf{x})\) and a parameterized unitary \(U(\mathbf{\theta}_i)\) or data-encoding unitaries and parameterized unitaries with trainable weights can be combined to a single unitary where the weights and biases are applied directly to the inputs to compute the effective rotation angles for the unitary:
\begin{align*}
L(k) = U(\mathbf{\theta}_k)U(x) &= U(\theta_k^{(0)},\theta_k^{(1)},\theta_k^{(2)})U(x^{(0)},x^{(1)},x^{(2)})\\
&= R(\theta_k^{(0)},\theta_k^{(1)},\theta_k^{(2)})R(x^{(0)},x^{(1)},x^{(2)})
\end{align*}
or
\begin{align*}
L(k) &= U(\mathbf{\theta}_k\circ \mathbf{x} + b_k)\\
&=R\left(\theta_k^{(0)} x^{(0)} + b^{(0)}, \theta_k^{(1)} x^{(1)} + b^{(1)},\theta_k^{(2)} x^{(2)}+ b^{(2)}\right)
\end{align*}
The quantum circuits for both alternatives are shown in Figure~\ref{fig:data_reuploading}. In both cases, the unitaries $U$ are typically chosen as arbitrary single qubit rotations $U=R(\phi,\theta,\omega)=R_Z(\omega)R_Y(\theta)R_Z(\phi)$ defined by three rotation angels (around the $z$- and $y$-axis).

The output is measured, and the results are used to adjust the parameters \(\mathbf{\theta}\) through classical optimization algorithms, aiming to minimize a predefined loss function. In the case of just two classes, we can simply measure in the computational basis to obtain the probabilities for $\ket0$  and $\ket{1}$ which are assigned to the first class and the second class, respectively. In this case, the vector of probabilities can just be compared to a ground truth vector using a standard loss function from classical machine learning (e.g., binary cross entropy or mean squared error).

In the case of more than two classes, a specific target state $\ket{\tilde\psi}_{C}$ must be specified for every class $C$ and the individual target states need to be as different from each other as possible. In terms of their vector representation, target state vectors need to be \emph{approximately} orthogonal to each other.

The classification is then performed by measuring how close a particular sample is to each of the target states, which leads to the concept of fidelity. The fidelity is a metric that measures how close two quantum states are. If both states are identical, the fidelity equals one and if they are orthogonal it is zero. 

To calculate the loss of our re-uploading classifier, we want to calculate the fidelity between the correct target state $\ket{\psi}_{C}$ and the state $\ket{\psi}_\mathbf{x}:=U(\mathbf{\theta}, \mathbf{x})\ket{0} $ of the data point $\mathbf{x}$. The fidelity $F=F(\ket{\psi}_{C},\ket{\psi}_\mathbf{x})$ can be computed by estimating the probability of measuring $\ket{0}$ for the state $\ket{\psi}_F = U_{C}^{\dagger}U(\mathbf{\theta}, \mathbf{x})\ket{0}$ where $U_{C}^{\dagger}$ denotes the inverse of the unitary that prepares $\ket{\psi}_{C}$. By summing up the fidelities for $M$ data samples $\lbrace\mathbf{x}_m\rbrace_{m=1}^M$ and their corresponding ground truth classes $\lbrace y_m\rbrace_{m=1}^M$, we obtain the loss $\chi_F^2$.
\begin{equation*}
\chi_F^2(\theta)=\sum_{m=1}^M\left(1-F(\ket{\psi}_{C=y_m},\ket{\psi}_{\mathbf{x}_m}))\right)
\end{equation*}

\begin{figure}[t!]
\centering
\Qcircuit @C=0.85em @R=.9em  @!R{
\lstick{|0\rangle} & \qw & \gate{L_{1}(1)} & \ctrl{1} & \gate{L_{1}(2)} & \ctrl{1} & \qw & \cdots & & \ctrl{1} & \gate{L_{1}(N)} & \qw  \\
\lstick{|0\rangle} & \qw & \gate{L_{2}(1)} & \ctrl{-1} & \gate{L_{2}(2)} & \ctrl{-1} & \qw & \cdots & & \ctrl{-1} & \gate{L_{2}(N)} & \qw \\
\vspace{0cm}
\gategroup{1}{3}{2}{4}{.7em}{--}
\gategroup{1}{5}{2}{6}{.7em}{--}
\gategroup{1}{11}{2}{11}{.7em}{--}
}
\caption{Data re-uploading classifier quantum circuit ansatz for two qubits with entanglement. Adapted from \cite{perez-salinas_data_2020}.}
\label{fig:data_reuploading_two_qubit}
\end{figure}
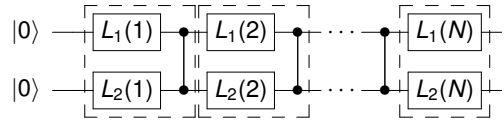
The data re-uploading classifier can also be trivially extended to work with two or more qubits as shown in Figure~\ref{fig:data_reuploading_two_qubit}. To that end, a layer of entangling gates such as CNOT or CZ can be added after each rotation gate layer such that the re-uploading classifier now resembles very much the generic PQC architecture presented in Figure~\ref{fig:ffnn_qnn}. In that case, the fidelity cost function described above is not needed any more but the classes $C_k$ can be directly mapped to measurement outcomes in the computational basis ($C_0\mapsto\ket{00}$, $C_1\mapsto\ket{01}$, $C_2\mapsto\ket{10}$, $C_3\mapsto\ket{11}$) and the cost can be computed directly based on the resulting probability distribution using the standard cross-entropy cost function (after suitable marginalization/renormalization of the probability scores if there are more measurement outcomes than classes).

\section{QC AND ML VISUALIZATION METAPHORS}
%\section{QC and ML Visualization Metaphors}
In the following section, we will provide a summary of the key visualization techniques available for quantum computing as well as for machine learning. 

\subsection{Quantum Computing}
\begin{figure*}[ht]
     \begin{subfigure}[b]{0.49\textwidth}
         \centering
         \raisebox{0.25in}{\includegraphics[width=\textwidth]{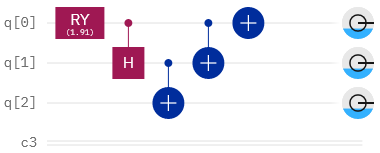}}
         \caption{Circuit for the preparation of a $W$-state $\ket{W}=\frac{1}{\sqrt{3}}(\ket{001}+\ket{010}+\ket{100})$.}
         \label{fig:w_state_circuit}
     \end{subfigure}
     \hfill
     \begin{subfigure}[b]{0.45\textwidth}
         \centering
         \includegraphics[width=0.9\textwidth]{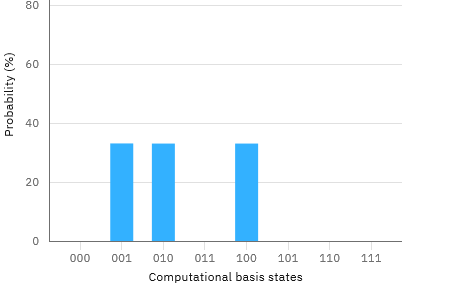}
         \caption{Probability distribution of basis states for the $W$-state}
        \label{fig:w_state_histogram2}
     \end{subfigure}
     \label{fig:circuits_state_distribution}
     \caption{Illustration of a quantum circuit and its corresponding basis state distribution after measurement.}
\end{figure*}

Compared to classical computing paradigms and programming languages, visualization is already integrated at the very core of writing down a quantum algorithm: Even though it can be done using a series of instructions or equations that were introduced in the previous section, the standard way to write down a quantum algorithm is in the form of quantum circuit diagrams, which provide an overview of the global structure of the algorithm as well as its local details. An example is shown in Figure~\ref{fig:w_state_circuit}.
This notation closely resembles the way music sheets are written but despite the great overview that quantum circuit diagrams provide one problem remains: Most readers of music will not know how the music sounds before they play it and this also holds for the execution of quantum circuits and their resulting or intermediate states.

Hence, we have to distinguish between the visual programming language of circuit diagrams or related concepts such as tensor network representations or ZX diagrams and methods to visualize quantum states, state spaces, or measurement results.

The most straightforward way to visualize a general $n$-qubit state is using a bar chart where the height of each bar corresponds to the magnitude of the complex amplitude $\alpha_k$ of its corresponding basis state $\ket{k}$.
The complex phase can simultaneously encoded in the color of the bar while bars with a negative phase can point downwards. This type of visualization is very helpful in understanding the ''inversion about the mean'' step used in Grover's algorithm but is otherwise generally hard to interpret.

Next, by taking the absolute square  of each amplitude $\vert\alpha_k\vert^2$, one obtains the probability mass function determining the distribution of measured basis states, which is essential for identifying the states that are most likely to be measured. See Figure~\ref{fig:w_state_histogram2} for an example. If the $\alpha_k$ are unknown,  this probability mass function must be approximated by using histogram plots of the measurement results.

In summary, these amplitude-based visualization techniques are certainly useful in understanding some algorithmic principles but there is also a loss of essential information in the process. In the following, we present some more advanced visualization methods that are also able to visualize the effect of specific gates, for example.

\subsubsection{Bloch Sphere}
The Bloch Sphere is a powerful visualization tool used in quantum computing to represent the state of a single qubit. It visualizes the state of a qubit on the surface of a sphere, where any point on the sphere represents a possible state of the qubit. This is particularly useful for understanding single-qubit operations such as rotations and the superposition principle.

The Bloch Sphere is a geometric representation of the single qubit states using the surface of a sphere.
In this representation, the north and south pole of the sphere corresponds to the state \(\ket{0}\) and \(\ket{1}\), respectively. Any point on the surface of the sphere represents exactly one possible state of the qubit, with points on the equator representing states of equal superposition (with various relative phases).

\begin{figure}[ht]
\centerline{\includegraphics[width=10pc]{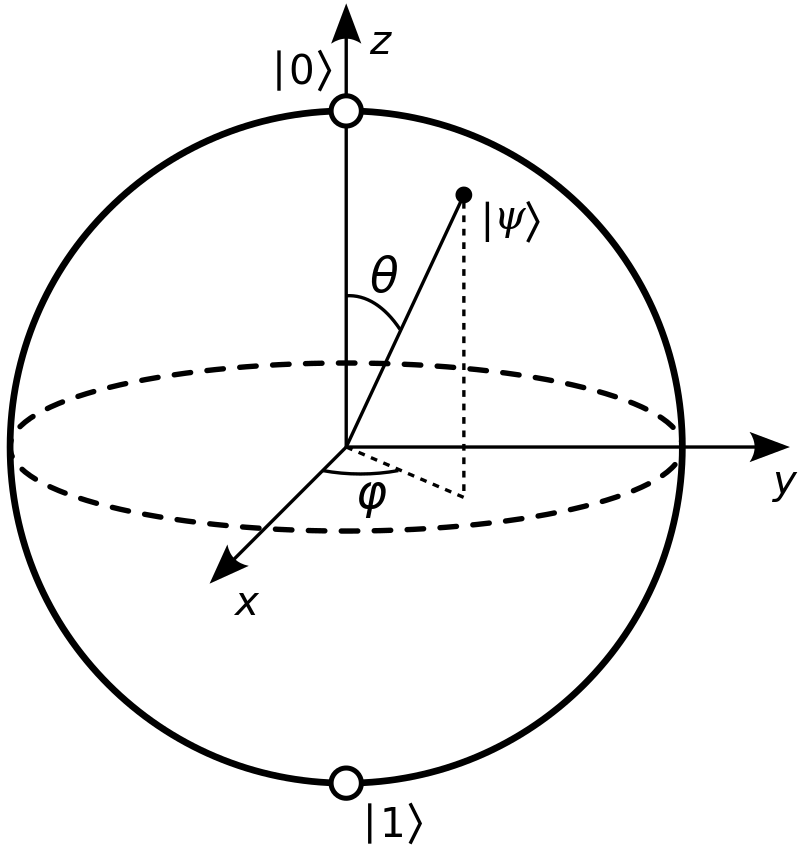}}
\caption{Illustration of the state $\ket{\psi}$ as a point on the Bloch sphere (Source: Wikimedia Commons).}\vspace*{-5pt}
\label{fig:bloch_sphere}
\end{figure}

A point on the Bloch Sphere (cf. Fig~\ref{fig:bloch_sphere}) can be specified using two angles $\theta$ and $\phi$. Using these angles, the state equation in Equation \eqref{eq:qubit_state} can be used to express the general state of a qubit as:
\[|\psi\rangle = \cos\left(\frac{\theta}{2}\right)|0\rangle + e^{i\phi}\sin\left(\frac{\theta}{2}\right)|1\rangle\]
Here, \(\theta\) ranges from 0 to \(\pi\) and represents the polar angle, while \(\phi\) ranges from 0 to \(2\pi\) and represents the azimuthal angle (the phase difference between the \(\ket{0}\) and \(\ket{1}\) components).

\begin{figure}[ht]
\centerline{\includegraphics[width=\linewidth]{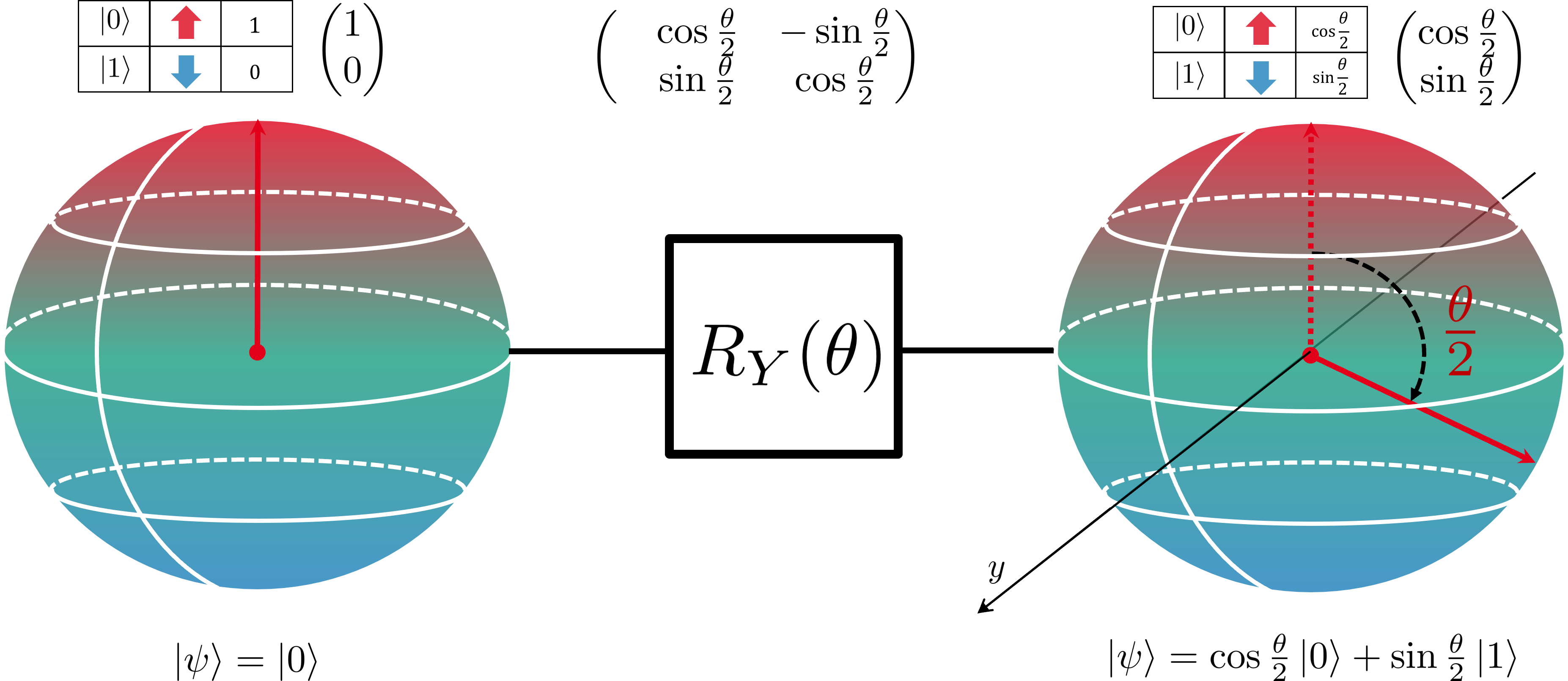}}
\caption{Illustration of the application of a $R_Y$ rotation gate using the Bloch Sphere.}
\label{fig:ry_rotation_gate}
\vspace*{-5pt}
\end{figure}

The Bloch Sphere is not just useful for visualizing qubit states; it also provides a way to visualize quantum operations (gates) as rotations on the sphere.
For example, the application of the $R_Y(\theta)$-gate corresponds to a rotation by $\frac{\theta}{2}$ around the $y$-axis, as shown in Figure~\ref{fig:ry_rotation_gate}.

While the Bloch Sphere elegantly represents single-qubit states and operations, it cannot directly visualize more complex phenomena like entanglement between multiple qubits, which require more sophisticated tools or higher-dimensional representations. One example is the so-called Q-Sphere.

\subsubsection{Q-Sphere}
The Q-sphere is a sophisticated visualization technique designed to represent the state of multi-qubit quantum systems, emphasizing the superpositions and entanglement that characterize quantum computing.
The Q-sphere can illustrate the states of multiple qubits simultaneously, providing a more comprehensive view of a quantum system's overall state.

The Q-sphere is a three-dimensional representation where each point on the sphere corresponds to a possible quantum state of the qubits being visualized. The key feature of the Q-sphere is its ability to represent both the amplitude and phase of each state in a multi-qubit system, as well as the entanglement between qubits.

\begin{figure}[ht]
\centerline{\includegraphics[width=15pc]{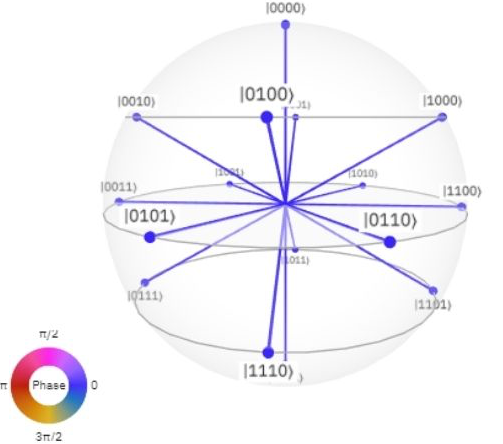}}
\caption{Q-Sphere visualization of a uniform superposition over four qubits.
The phase of the complex amplitude is color-coded.
Lattitude encodes the hamming weight of the respective basis state.}
\label{fig:q_sphere}
\vspace*{-5pt}
\end{figure}

An example of Q-sphere visualization for 4 qubits is shown in Figure~\ref{fig:q_sphere}.
Each quantum state of a multi-qubit system is represented by a dot or a vector on the sphere.
The position of the dot on the sphere reflects the state's quantum numbers (or basis states), with the north and south poles typically representing the all-zero (\(|00...0\rangle\)) and all-one (\(|11...1\rangle\)) states, respectively.
The latitudes of the other states are given by their Hamming weights.
The amplitude of a quantum state is represented by the size of the dot, while the color or shading of the dot indicates the phase of the state's amplitude.
This way, the Q-sphere visually encodes the complex probability amplitude of each basis state in the multi-qubit quantum system.
Superposition states are shown by the presence of multiple dots, indicating the system's quantum state is a combination of these basis states. Entangled states \emph{can} be visualized using the Q-sphere, however, it is often difficult to decide whether the shown state is separable or entangled just by looking at its Q-sphere visualization.

While the Q-sphere provides a powerful tool for visualizing complex quantum states, its utility diminishes as the number of qubits increases, due to the exponential growth in the number of possible states. For systems with many qubits, the Q-sphere can become cluttered, making it difficult to discern detailed information about the state.

\subsubsection{Q-Simplex}
For our algorithm visualization concept in the latter section we introduce a new visualization specifically for the case of two qubits which we call the \emph{Q-Simplex}. The biggest challenge when visualizing multi-qubit states is the exponential increase in the degrees of freedom: Already the description of a two-qubit state requires six degrees of freedom (eight for the state vector while one can be removed due to the normalization condition and another one due to irrelevance of the global phase)

\begin{figure}[ht]
\centerline{\includegraphics[width=15pc]{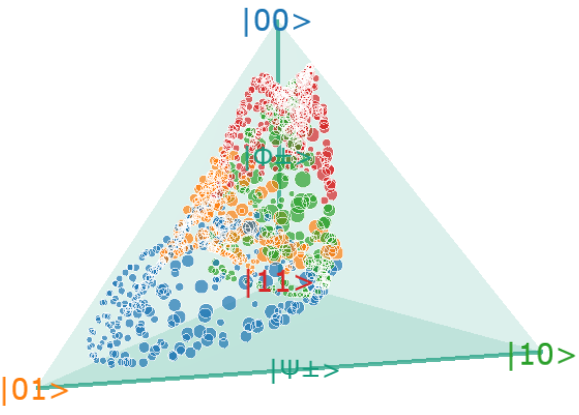}}
\caption{Q-Simplex visualization. The two edges containing the maximally entangled Bell states as midpoints are highlighted. In this example, marker color encodes a class label while the size corresponds to the concurrency $C$ of each state.}
\label{fig:2qsimplex}
\vspace*{-5pt}
\end{figure}

Our approach is based on the observation that the distribution of measurement probabilities in the computational basis for a general two-qubit state \eqref{eq:twoqubit_state}, together with the normalization constraint $|a|^2 + |b|^2 + |c|^2 + |d|^2 = 1$, can be interpreted as a (probability) 3-simplex (i.e., a solid tetrahedron in three-dimensional space). Hence, general two-qubit states can be mapped to points onto a transparent tetrahedron, where the vertices correspond to the four computational basis states ($\ket{00}$, $\ket{01}$, $\ket{10}$, $\ket{11}$). Furthermore, two of the edges can directly be associated with entangled states. The midpoints of these edges, in particular, correspond to the four bell states.
While this representation ignores the complex relative phases between the basis states, some of this information can be brought back into the visualization by encoding it in the color or size of the points. An example of this visualization is shown in Figure~\ref{fig:2qsimplex}.

\subsection{Machine Learning and Neural Networks}
\begin{figure*}[ht]
\centering
   \begin{subfigure}[t]{0.45\textwidth}
         \centering
         \raisebox{0.20in}{\includegraphics[width=\textwidth]{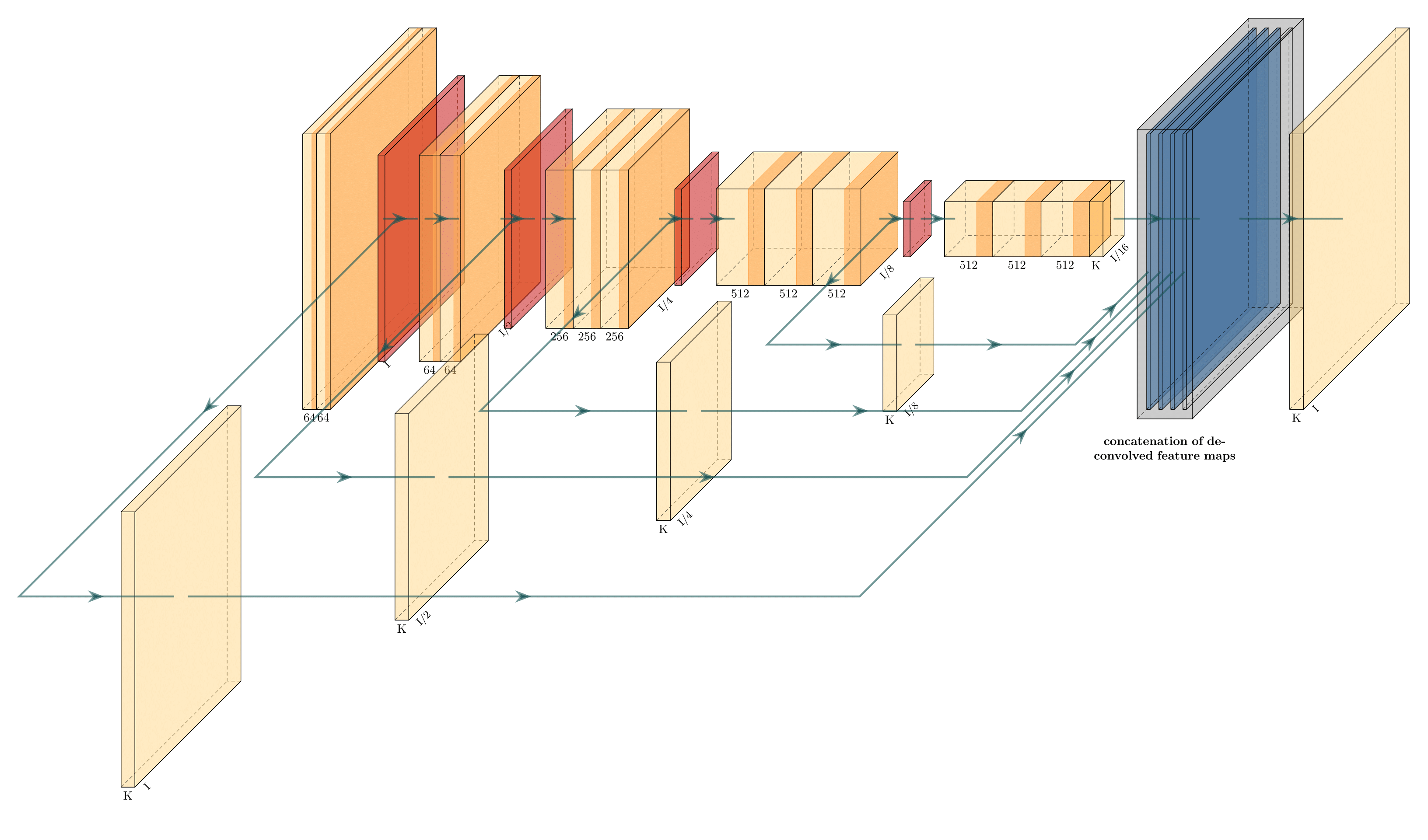}}
         \caption{Visualizing a complex neural network architecture. Source: \cite{iqbal_harisiqbal88plotneuralnet_2018}.}
         \label{fig:PlotNeuralNet}
     \end{subfigure}
     \hfill
    \begin{subfigure}[t]{0.45\textwidth}
         \centering
         \includegraphics[width=\textwidth]{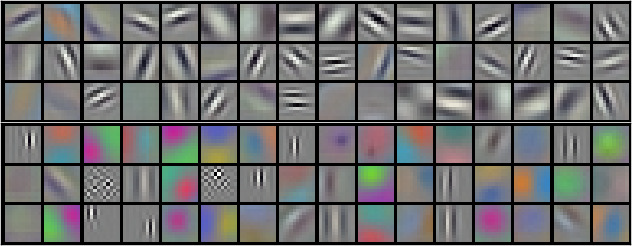}
         \caption{Visualizing filters of a CNN. Source: \cite{krizhevsky_imagenet_2012}.}
         \label{fig:cnn_filter_viz}
     \end{subfigure} 
       \begin{subfigure}[t]{0.50\textwidth}
         \centering   
            \raisebox{0.3in}{\includegraphics[width=0.75\textwidth]{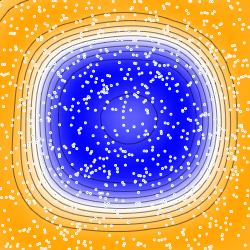}}
         \caption{Visualizing the decision boundary of a classifier using heat maps and contour lines.}
         \label{fig:decision_boundary}
     \end{subfigure}
     \hfill
    \begin{subfigure}[t]{0.45\textwidth}
         \centering
         \includegraphics[width=\textwidth]{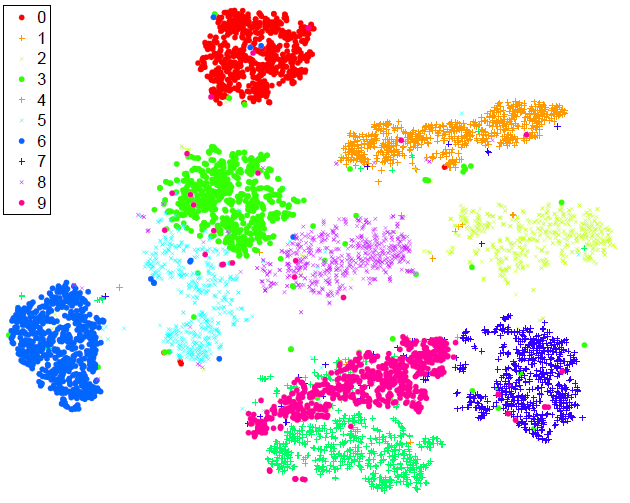}
         \caption{Visualizing the t-SNE algorithm for dimensionality reduction. Source: \cite{maaten_visualizing_2008}.}
         \label{fig:tsne_viz}
     \end{subfigure} 
     \vspace{8pt}
     \caption{Overview of different classical machine learning visualization techniques.}
     \label{fig:classical_ml_visualization}
\end{figure*}

Visualization techniques for neural networks are crucial for understanding, debugging, and interpreting the models' decision-making processes. In the following, we present a selection of different groups of visualization methods. Corresponding examples for the techniques are also shown in Figure~\ref{fig:classical_ml_visualization}.

\subsubsection{Architecture Diagrams}
The most elementary visualization for neural networks is just to display their structure (number of (hidden) layers, type of layers, number of neurons).
A very basic example of a simple feed-forward neural network is shown in Figure~\ref{fig:ffnn_qnn}.
Here, nodes represent the neurons, and weights are represented by edges between neurons, while biases are typically omitted.
However, with the increasing complexity of the neural network design space, more refined methods were developed that also incorporate visualizations of the internal state.

\subsubsection{Internal State Visualization}
This can be either done considering the neural network in isolation or while the neural network is processing a specific input.
In the first case, the values or magnitudes of the neural network's weights are visualized either via the thickness or color of the edges or as a heatmap considering the weights matrix of the whole layer. 

In the second case, one considers the values at the nodes before and after the application of the activation function. For the special case of convolutional neural networks (CNN), one can visualize the filters of the convolutional layers to see the patterns or features that the network is specifically looking to detect.
This is especially useful in the early layers of a CNN, where filters might be looking for simple edges or colors.
Instead of displaying a heat map, the activations of a whole layer can be considered as points in high-dimensional latent feature space (with the number of neurons in the layer being equal to the dimension of this space). 

Dimensionality reduction techniques like t-SNE (t-Distributed Stochastic Neighbor Embedding, \cite{maaten_visualizing_2008}) and PCA (Principal Component Analysis, \cite{pearson_liii_1901}) can be used to visualize high-dimensional data in two or three dimensions.
This can be particularly useful for understanding the clustering and separation of features in the latent space.

\subsubsection{Decision Boundaries}
Decision boundaries visualize how a neural network classifies different regions of the input space, which is useful for understanding the complexity and flexibility of the model in separating classes.
This is, however, only possible when the input space is restricted to two dimensions.
For high-dimensional input spaces, the decision boundary is merely an abstract concept or needs to be combined with dimensionality reduction techniques.
It is however also instructive from a theory point of view because one of the goals that deep neural networks with non-linear activation functions (and also classical as well as quantum kernel methods) try to achieve is a linear decision boundary in the latent, high-dimensional feature space.

\subsubsection{Training Diagnostics}
Training diagnostics refers to the visual representation of the (performance) metrics during the training process.
For example, this includes monitoring the loss values or accuracy metrics on the test and validation set and presenting them as a line plot with the number of epochs on the $x$-axis.
This provides insights into the optimization process and how easily the network can converge to a solution, or if the model overfits.

\section{ALGORITHM VISUALIZATION CONCEPT}
%\section{Algorithm Visualization and Interactivity Concept}
For our visualization concept of a quantum machine learning model, we focus on what can be learned from the TensorFlow Playground application. 

 \begin{figure*}
 \centerline{\includegraphics[width=\linewidth]{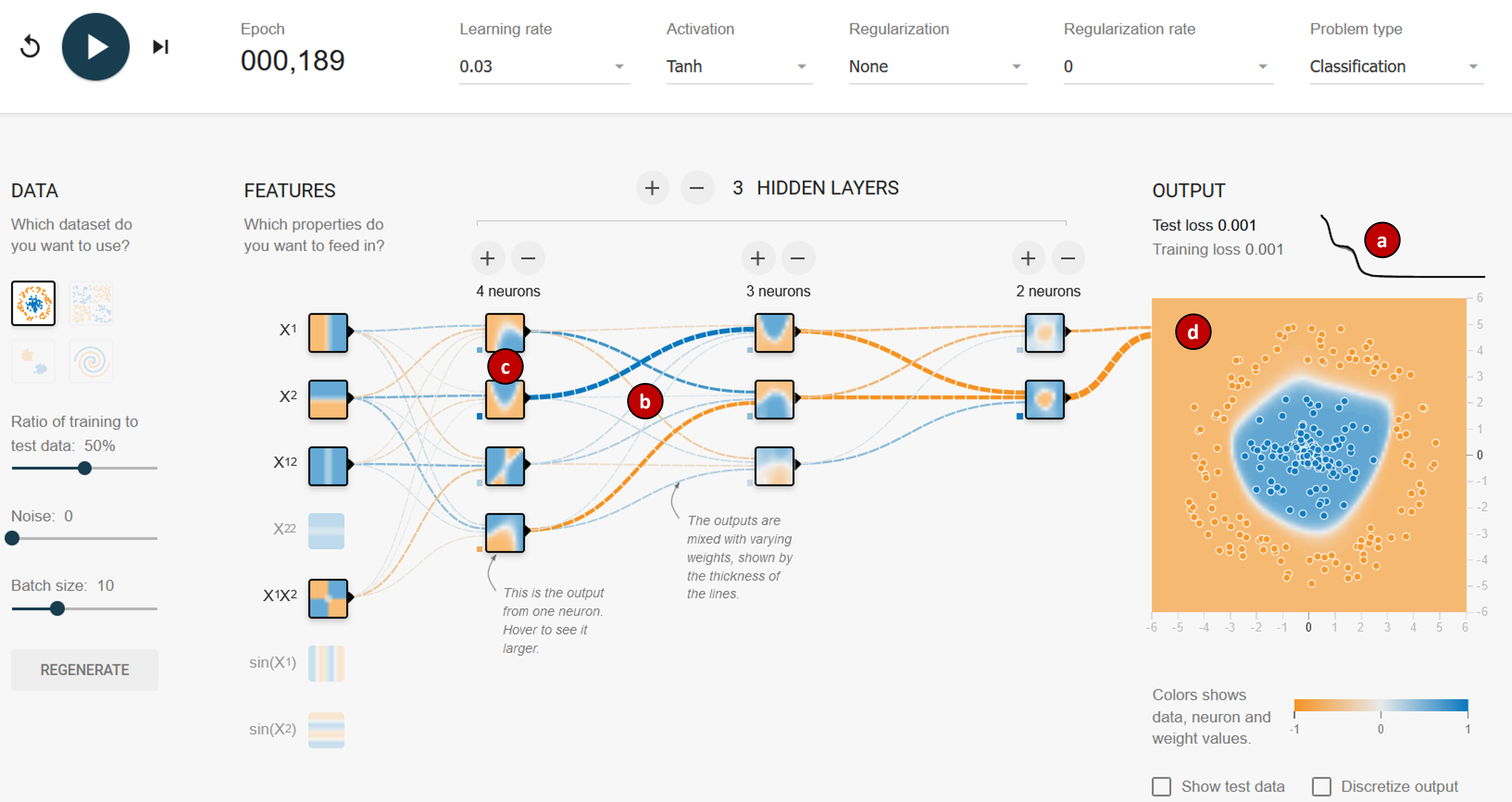}}
 \caption{Screenshot of the Tensorflow Playground web application highlighting different visualization techniques: (a) Accuracy over the training process, (b) weights visualization, (c) decision boundary heat maps on neuron level, (d) final decision boundary heat map. Source: \url{https://playground.tensorflow.org}.}\vspace*{-5pt}
 \label{fig:tf_playground_annotated}
 \end{figure*}

\subsection{TensorFlow Playground}
TensorFlow Playground is an interactive web application that visualizes the learning process of a neural network in real-time, making the abstract concepts of neural network operations more tangible. 
Users can configure the number of hidden layers and the number of neurons in each layer, visually constructing the network’s architecture.
The choice of activation function (e.g., ReLU, Sigmoid, Tanh) for each layer can be selected, affecting how neurons process inputs.
Users can also choose from several datasets (with different levels of complexity) and task types (classification or regression).
The playground visually represents these datasets, allowing users to see how the network's performance varies with different types of data.

As the training progresses, TensorFlow Playground updates the visualization of the decision boundary or regression line in real time, showing how the model improves its predictions over iterations.
Users can adjust the learning rate and observe how quickly or slowly the model learns, as well as monitor the number of epochs (iterations over the dataset) the model has trained for.
Users can experiment with different regularization methods (L1, L2, and dropout) and parameters, observing their effects on the learning process and how they help prevent over-fitting.
Adjustments to the learning rate and other hyperparameters allow users to see how these factors influence the speed and stability of learning. The immediate visual feedback on adjustments made to the network configuration or training parameters enables users to experiment with and understand the effects of their choices on the learning process.

In terms of visualization, most of the techniques that are described in the above section are employed in TensorFlow Playground. Figure~\ref{fig:tf_playground_annotated} shows an annotated overview of the methods. The application displays a graph of the loss function over time (a), illustrating how the model’s performance improves (or does not) as training progresses.
This graph is key for understanding the optimization process and how the model minimizes its error.
Next, as the training progresses sign and magnitude of the weights (b) are displayed via the color and the thickness of the edges connecting the neurons.
For every neuron a small decision boundary heat map (c) is displayed that indicates how classification results based on the sign of this neuron activation value would look like.
Lastly, a global decision boundary heat map (d) for the full model is displayed together with the original data set next to the output neurons.

\subsection{Concept for a Quantum Machine Learning Playground}
To come up with a similar concept for the even more challenging and abstract field of quantum machine learning, three main points must be addressed:
First, the choice of a concrete quantum machine learning algorithm that is suitable for visualization.
Second, the design of the graphical user interface (GUI) which also determines the degree of interactivity, and finally the choice of visualization techniques and metaphors that support the overall educational goal of creating a quantum machine learning playground.

\subsubsection{Algorithm Choice}
Finding a suitable algorithm that is simple enough that it can be easily visualized, yet still relevant and representative for a larger class of similar quantum machine algorithms is challenging.
An important class of algorithms in the era of noisy intermediate scale quantum (NISQ) devices are parameterized quantum circuits (PQC) with an architecture similar to the one presented in Figure~\ref{fig:ffnn_qnn}.
The architecture presented here can be modified in a variety of ways to account for different encoding methods and is also relevant for other algorithm classes such as kernel methods, in particular quantum kernel alignment methods.
Furthermore, also the structural similarity presented in Figure~\ref{fig:ffnn_qnn} helps people with a background in machine learning but not necessarily in quantum computing get a grasp of the method faster by seeing the correspondence between the known and the new elements.

On the other hand, what makes every PQC (or, in general, all quantum algorithms) difficult to visualize is the entangled multi-qubit states that are involved.
As discussed in the theory background section, a characteristic feature of entangled states is that they cannot be described locally, using the individual states of their single-qubit constituents, anymore.
This is in strong contrast to what is possible with individual neurons in the classical realm.
Of course, the whole state can be globally described at any layer as a vector of $2^n$ complex numbers, but this does not solve the non-locality problem and has the added challenge of visualizing complex amplitudes rather than just real values.

A suitable choice, that can address or at least alleviate these issues, is the data re-uploading classifier that was introduced in the theory background section which combines several very desirable properties:
The most important property is a corresponding universal approximation theorem (UAT) that was proven for the algorithm and which states that any continuous, real-valued function can be arbitrarily well approximated by even a single-qubit data re-uploading quantum model when the number of layers is increased accordingly.
In some sense, the data re-uploading can trade off the length versus the width (i.e., the number of qubits) of the quantum circuit necessary for its implementation.
Of course, from a theoretical as well as hardware implementation perspective, no quantum advantage can be expected here because few-qubit circuits can be efficiently simulated classically, which is also done in our implementation, to make a responsive graphical user interface possible.

From a visualization perspective, this enables us to visualize the full state space of the algorithm in the single-qubit case using the Bloch sphere as a well-known and established visualization concept from textbooks, while still having the guarantee that the algorithm is a universal classifier. At the same time, using the Q-simplex visualization, we are also able to explore the two-qubit case which includes the concept of entanglement as one of the key resources for quantum computing. As a result the algorithm also has a closer resemblance with the generic architecture of PQC-based models and is a building block for advanced data encoding methods as argued in \cite{schuld_effect_2021}.

\subsubsection{GUI Design}
The design of the GUI is rather straightforward due to the structural similarity to a typical machine learning model and training workflow. There are buttons and controls for starting and stopping the training process, as well as controls for selecting a data set, architecture, and (hyper) parameters for the training process.
Based on the capabilities of the underlying plotting library, most of the visualizations can be zoomed and rotated, and hover overlays display additional information.

\subsubsection{Visualization Techniques}
From the classical machine learning techniques, we use line plots for the training metrics, such as accuracy on the train and test set and the loss function values.
The data sets are displayed as scatter plots with different colors for different classes.
Furthermore, we also use the decision boundary heat maps augmented by two additional scatter plots, that also show the ground-truth decision boundaries and a different color code for correctly and wrongly classified data samples.

For the very core of our quantum machine learning visualization, we use the Bloch sphere and Q-simplex representation, which we augment with additional information such as the target state.
As briefly discussed in the previous section, the Bloch sphere has the advantage that it is not only a state visualization but also a visualization of the whole state space.

Consequently, not just a single point can be displayed after embedding as a quantum state but the whole data set can be displayed simultaneously on a single Bloch sphere. 
For the Q-simplex this is also true with the restriction that the full quantum states are projected to their measurement distributions such that some information is lost in the process. Nevertheless, it is more important to visualize how states move towards their target states rather than display full relative phase information.

Following the layer structure as shown in Figure~\ref{fig:data_reuploading}, we can display a Bloch sphere (or Q-simplex, respectively) after each layer of the model mimicking the neural network representation in TensorFlow playground.
That way, the effect of every parameterized rotation can be observed, as the input data set proceeds towards the output measurement layer of the model.
At the same time, it can be observed how the points of the different classes are separated and how they are moved towards their respective target states on the sphere (or simplex).

%\section{Implementation Details}
\section{IMPLEMENTATION DETAILS}
 \begin{figure*}[ht]
\includegraphics[width=\linewidth]{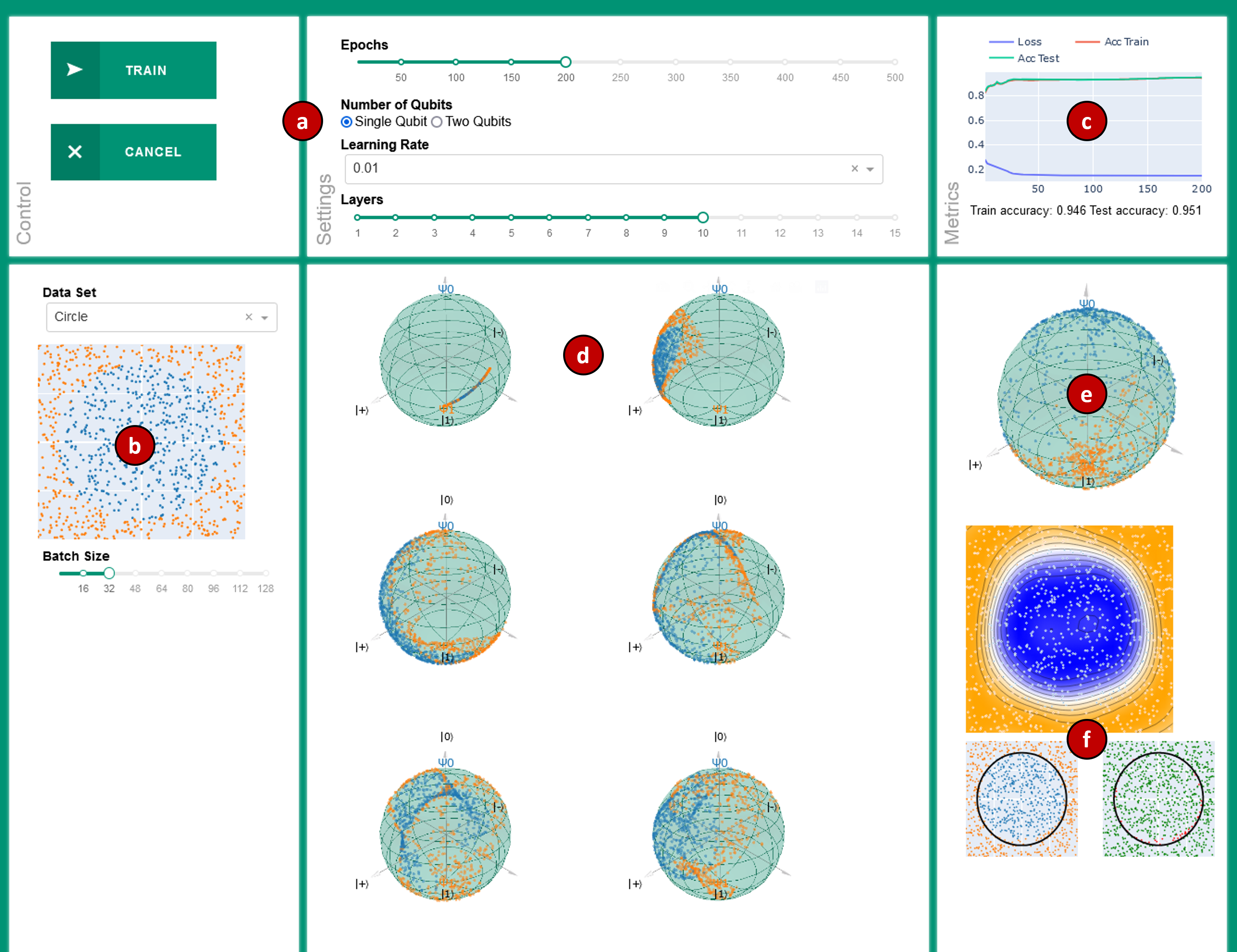}
 \caption{Screenshot of the QML Playground App highlighting different GUI and visualization elements: (a) training control elements, (b) dataset display and controls, (c) line plots of accuracies over the training process, (d) Bloch sphere visualization of the data set across the different layer, (e) final Bloch sphere visualization, (f) decision boundary heat map and data set with ground truth decision boundary.}\vspace*{-5pt}
 \label{fig:qml_playground_annotated}
 \end{figure*}
\begin{figure}[ht]
\includegraphics[width=15pc]{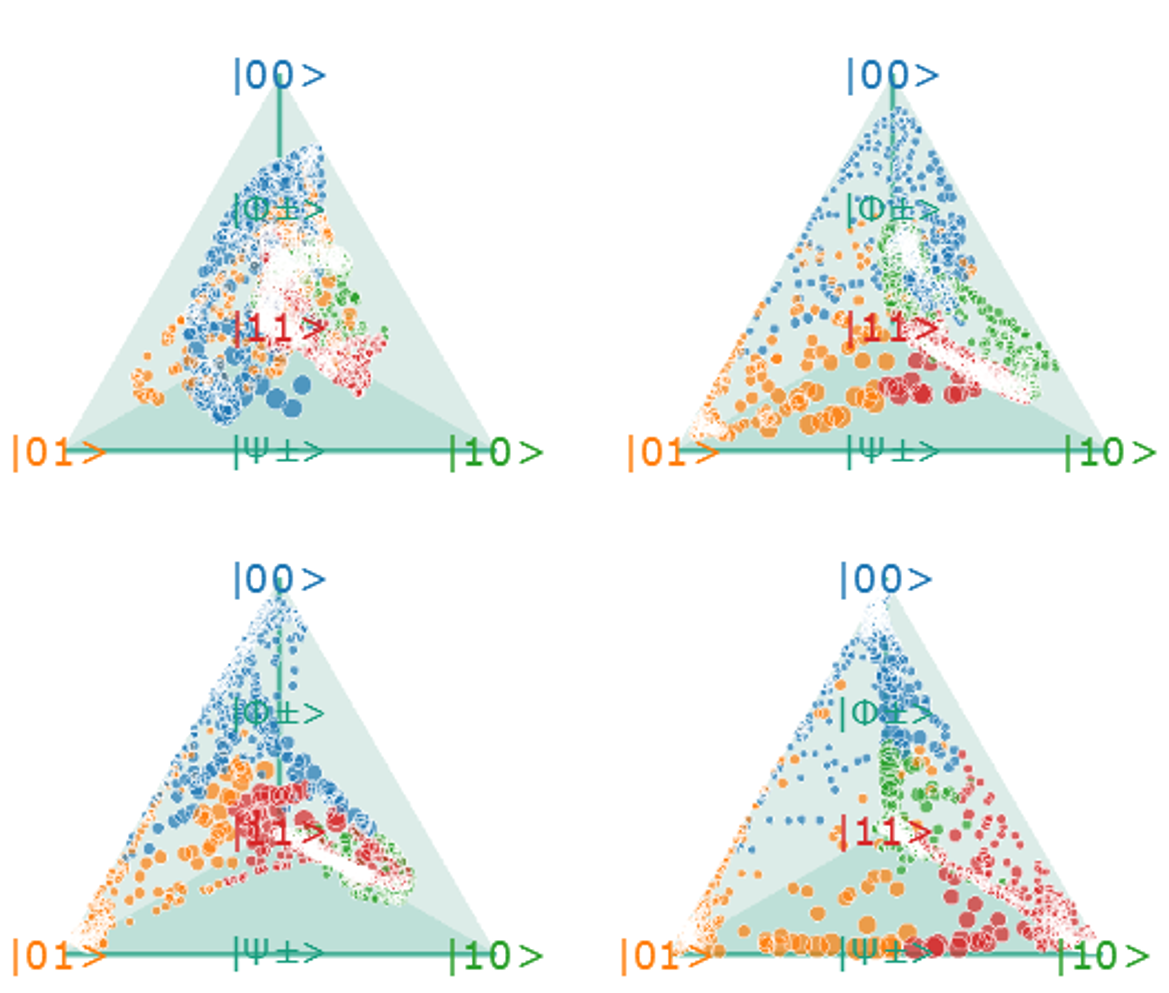}
 \caption{Section of the main panel displaying the states after the last four layers of a two-qubit reuploading model with 10 layers and a classification task for 4 classes.}\vspace*{-5pt}
 \label{fig:2qsimplex_models}
 \end{figure}

The GUI of the current implementation is shown in Figure~\ref{fig:qml_playground_annotated}.
The GUI consists of (a) training controls and (b) data set display with additional controls for data set selection and batch size.
On the upper right is a (c) line plot for the important training metrics.
The (d) main panel shows the main visualization of the embedded data set after every layer of the model. The (e) final state before measurement of the state is displayed in a dedicated panel together with (f) decision boundary heat maps.
In Figure~\ref{fig:2qsimplex_models} we also show a section of the two-qubit version of the main panel (d).

The web application is written in Python, using the Dash \cite{manley_dash_2023} and Plotly libraries for the frontend and PyTorch \cite{paszke_pytorch_2019} for the backend implementation of a state vector simulator for the data-reuploading algorithm.
Optimization is performed using the Adam optimizer and in contrast to what would be possible on real quantum hardware, we compute the gradients using PyTorch's built-in automatic differentiation during the forward pass of a whole mini-batch of data points, after which the weights are updated using backpropagation.
Furthermore, the intermediate state is extracted after each layer, such that it can be visualized in the main panel of the application. 

In an earlier version of the implementation standard quantum development libraries such as Qiskit \cite{aleksandrowicz_qiskit_2019} or Pennylane \cite{bergholm_pennylane_2022} were used.
However, the training process was usually too long for a responsive and interactive web application, since these libraries are typically optimized for the computation of circuits with many qubits rather than a lot of long circuits with just a single qubit. The application will be made available upon publication in a public code repository\footnote{\url{https://github.com/Fraunhofer-AISEC/qml-playground}}.

%\section{Directions for Future Work}
\section{DIRECTIONS FOR FUTURE WORK}
The visualization concept and implementation presented in the previous section is merely a first step towards effective visualization of quantum machine learning concepts that cover the basics of its classical template. 
Considering just the implementation and the GUI some minor features are easy to add, such as adding noise to the data set or adding standard regularization techniques to the training process.

When it comes to additional visualization techniques, the weights and biases of the individual layers could also be visualized using heat maps in the simplest case. However, it might be worthwhile to consider the special meaning of these biases and weights being rotation angles (or angle multipliers) around three different axes. Some specialized techniques could be employed for the visualization of rotations (or so-called manifold-valued data) in general.

Furthermore, it was recently shown that there are some intriguing connections between the capabilities of a quantum machine learning model and the spectrum of its underlying quantum circuit when the latter is interpreted as a Fourier sum. Building meaningful visualizations from these quantities might be very instructive in understanding how to build a good model.

Finally, the biggest challenge in visualizing quantum machine learning models in particular, and quantum algorithms, in general, is to find a scalable approach for entangled multi-qubit states. 
The Q-simplex visualization we suggested is a first approach towards this goal but by design cannot be extended easily beyond the 2-qubit cases. Nevertheless, maybe different three-dimensional objects that can be expressed as convex combinations of a larger number of variables might be promising candidates.

\section{CONCLUSION}
Inspired by the success of classical machine learning visualization tools, we developed a similar visualization tool for QML models. To that end, we provided an overview of common visualization metaphors in machine learning and quantum computing and combined them into an interactive visualization concept for QML.  To make this concept more tangible we also discussed our concrete implementation for the data re-uploading universal quantum classifier. This classifier was chosen because it is representative of a large class of QML models, particularly relevant for the data-encoding problem in QML in general, and yet possesses properties that make it very suitable for visualization. The interactive application we have implemented can help to understand QML algorithms through hands-on, visual exploration and make it more accessible for newcomers to this field. 

While our approach is a first step towards visualizing and understanding the challenges of QML, it also opens the door to numerous opportunities for future research. Several were discussed in the previous section. Among them, the meaningful and application-specific visualization of entangled multi-qubit states is certainly the most relevant one.

The evolution of quantum computing demands ongoing development of educational and research tools to match its pace. Future efforts should aim at expanding the range of quantum algorithms that can be visualized by combining existing visualization metaphors or creating new ones.

\section{ACKNOWLEDGMENTS}
The project/research is supported by the Bavarian Ministry of Economic Affairs, Regional Development and Energy with funds from the Hightech Agenda Bayern.

\bibliographystyle{IEEEtran}
\bibliography{references}

\begin{IEEEbiography}{Pascal Debus}{\,} is a research group leader at the Fraunhofer Institute for Applied and Integrated Security (AISEC). His research interests include topics at the intersection of quantum computing, machine learning, optimization, and IT security. He received an M.Sc. degree in physics from ETH Zurich. Contact him at pascal.debus@aisec.fraunhofer.de.
\vspace*{8pt}
\end{IEEEbiography}

\begin{IEEEbiography}{Sebastian Issel}{\,} is a researcher at the Fraunhofer Institute for Applied and Integrated Security (AISEC). His research interests are quantum computing, in particular applications of the quantum singular value transformation and optimization. He graduated from RWTH Aachen with a M.Sc. in Computer Science. Contact him at sebastian.issel@aisec.fraunhofer.de.\vspace*{8pt}
\end{IEEEbiography}

\begin{IEEEbiography}{Kilian Tscharke} {\,} is a researcher at the Fraunhofer Institute for Applied and Integrated Security (AISEC). His research interests are quantum computing, machine learning, IT security, and all possible combinations thereof. He graduated from FAU Erlangen-Nürnberg with a M.Sc. in Materials Science. Contact him at kilian.tscharke@aisec.fraunhofer.de.
\end{IEEEbiography}

%All biographies are limited to one paragraph, following the structure given here: each author's current role %and institution (to match the first page of the article); three to five current research interests; highest degree, %topic, and awarding institution (do not include year); professional memberships, such as the IEEE Computer %Society and any grade information; and contact information in the form of an email address.

\end{document}